\RequirePackage{lineno}
\documentclass[linenumbers,author-year,preprint,showkeys,preprintnumbers,amsmath,amssymb,superscriptaddress]{revtex4}

\usepackage{graphicx}
\usepackage{graphics}
\usepackage{subfigure}
\usepackage{dcolumn}
\usepackage{bm}
\usepackage{txfonts}
\usepackage{enumerate}
\usepackage{amsmath,empheq,mathrsfs}

\DeclareMathOperator{\sign}{sign}

\begin{document}


\centerline{Opinion formation on dynamic networks: identifying conditions for the emergence of partisan echo chambers}

{\small {\vskip 12pt \centerline{Tucker Evans$^{1}$, Feng Fu$^{1,2}$}

\begin{center}
$^1$ Department of Mathematics, Dartmouth College, Hanover, NH 03755, USA\\
$^2$ Department of Biomedical Data Science, Geisel School of Medicine at Dartmouth, Lebanon, NH 03756, USA
\end{center}
}}

\vskip 30pt

\begin{minipage}{142mm}
\begin{flushleft}

{\textbf{Running title:}}\, Modeling echo chambers
{\textbf{Manuscript information:}}\, Type of article: Research Article; Number of words in the abstract: $200$; Number of words in the main text: $5020$ (excluding abstracts, references, and figure legends); Number of figures: 4 main figures.\\
{\textbf{Author contributions:}}\, T.E. conceived the model, performed simulations and data analyses; T.E. \& F.F. constructed the application of the model to real data; T.E. \& F.F. wrote the manuscript. All authors gave final approval for publication.\\
{\textbf{Corresponding authors:}} \\
Tucker Evans \& Feng Fu\\
\small {Department of Mathematics, Dartmouth College\\
27 N. Main Street, 6188 Kemeny Hall\\
Hanover, NH 03755 USA\\
Email: Tucker.E.Evans.19@dartmouth.edu (T.E.); fufeng@gamil.com (F.F.)\\
Tel: +1 (603) 646 2293\\
Fax: +1 (603) 646 1312\\}
\end{flushleft}
\end{minipage}

\clearpage

\begin{center}
{
\begin{minipage}{142mm}
{\bf Abstract:} \, Modern political interaction is characterized by strong partisanship and a lack of interest in information sharing and agreement across party lines. It remains largely unclear how such partisan echo chambers arise and how they coevolve with opinion formation. Here we explore the emergence of these structures through the lens of coevolutionary games. In our model, the payoff of an individual is determined jointly by the magnitude of their opinion, their degree of conformity with their social neighbors, and the benefit of having social connections. Each individual can simultaneously adjust their opinion as well as the weights of their social connections. We present and validate the conditions for the emergence of partisan echo chambers, characterizing the transition from cohesive communities with consensus to divisive networks with splitting opinions. Moreover, we apply our model to voting records of the United States House of Representatives over a timespan of decades in order to understand the influence of underlying psychological and social factors on increasing partisanship in recent years. Our work helps elucidate how the division of today has come to be and how cohesion and unity could otherwise be attained on a variety of political and social issues. 

\vspace*{1\baselineskip}

\end{minipage}
}
\end{center}

\clearpage


\section{Introduction}

In a political context of growing partisanship and echo chamber formation, a natural question is how such a culture of division arose and what might be to come in the future if the trend continues. In order to address this question using an opinion formation model, it is necessary to take into account both the underlying motives of an individual's actions and also his or her position in the socio-political milieu. This environment consists of all the influences on that individual: his or her sources of information and the social pressures that he or she feels to conform to one view or another. Network science offers the theoretical basis for the quantification and description of these factors, and provides metrics for the modular structure implied by ``echo chambers'' and partisan politics. 

Over the past decades, making use of the extensive available data, researchers have been able to identify some common topological characteristics of social networks. For example, social networks are frequently found to be highly heterogenous with the small-world property~\cite{Albert:2000,Milgram:1967}. In the broader biological context, high degree nodes have been shown to play an essential role in promoting network robustness and increasing the efficiency of communication and flow through the network~\cite{Borgatti:2003,Albert:2000}. Furthermore, biological networks frequently exhibit organization into clustered communities of nodes (often referred to as modules), the recognition of which is in itself a topic of study~\cite{Newman:2003,Newman:2006,Mas:2010}. This line of research is in parallel with more theoretical considerations of how large networks can be categorized and described \cite{Zadorozhnyi:2012,Jacome:2010,Dorogovtsev:2008,Dorogovtsev:2002,Erdos:1959,Erdos:1960}. 

The discovery of these network topological features begs the question of how they influence the dynamics taking place on the network. Concerning opinion dynamics specifically, how does the network structure of a group of individuals change the way in which those individuals develop their own beliefs? Previous studies of opinion dynamics have explored population structures with the general characteristics of social networks with an emphasis on refining the opinion updating procedures employed by individual nodes in the network~\cite{sznajd2000opinion,deffuant2002can,sznajd2005sznajd,watts2007influentials,szabo2002three,sood2008voter,redner2017dynamics, jalili2017information,Friedkin:2016,Axelrod:1997}. 

Most recently, the coevolution of opinion dynamics and individual social connections has received increasing attention~\cite{holme2006nonequilibrium,zanette2006opinion,nardini2008s,Fu:2008}. In these models, Individuals modify their interactions with those who chronically disagree with them, usually with the purpose of avoiding discussion of the topics which arouse contention. Building on this prior work, we investigate opinion formation on dynamic social network through the lens of a coevolutionary game~\cite{perc2010coevolutionary}. The relevant structural information of the network includes the opinions of the individuals in the network and the relative influence that each individual exercises on his or her neighbors, concepts which can be understood as values attached to the nodes and to the edges, respectively. This paper will consider how individuals' opinions change alongside the structure of the network in which they are embedded, taking as a basic assumption that the two are fundamentally linked and that they mutually influence each others' evolution. Development of this kind can lead to complex structure within the network, an excellent example of which are the idealistically divided sub-communities that develop in the political sphere, from democrats and republicans~\cite{Polarization:2014,Partisanship:2016} to pro-vaccination and anti-vaccination groups~\cite{Schmidt:2018}. 

Improving our understanding of opinion evolution is essential in the context of modern politics. Partisanship has come to dominate the political sphere, both amongst the political elite and the general population\cite{Andris:2015,Conflict:2017,Polarization:2014,Partisanship:2016,Election:2016}, stalling political consensus on issues that demand rapid decision-making. It has become a major research concern to effectively understand circumstances that will lead to convergence of opinion and others that will lead to divergence of opinion and a weakening of information transfer~\cite{Antonopoulos:2017}. The model and methods provided in this work allow for the analysis of the current state of politics and the partial elucidation of how the division of today has came to be. 

\section{Model}

Let us denote by $\mathbf{X}$ the vector of opinions of all nodes in the network, where $X_{i}$ refers to the belief of the $i^{th}$ node. The values of the $X_{i}$ are allowed to range in the interval $[-1,1]$, where negative values indicate disagreement and positive values denote agreement with a given proposition. Greater magnitudes indicate stronger opinions. Let $\mathbf{C}$ represent the weighted adjacency matrix of the network, composed of the weights on the connections between any pair of two nodes in the network, with $C_{ij}$ referring to the connection between the $i^{th}$ node and the $j^{th}$ node. For the purpose of our model, the connections between nodes are considered to be symmetrical, meaning that $C_{ij} = C_{ji}$. The value of $C_{ij}$ can range between zero and one, where zero indicates that there is no connection between $i$ and $j$ and one indicates the strongest possible tie between the two. 

The payoff (= fitness) of a node in the network is determined as the sum of three different factors: the strength of the node's opinion, the node's level of agreement with neighboring nodes in the network (nodes with which the node in question has a non-zero connection), and the degree of the node (the total sum of the weights of the node's connections). The importance of each of these three components is described by the parameter $\alpha$, $\beta$, and $\gamma$, respectively.

For a network of size $N$, the fitness of an individual node $i$ is given by:
\begin{equation}
\label{eq:nodeFitness}
f_{i}(X,C) = \alpha|X_{i}|+\beta\sum_{i \neq j}C_{i,j}X_{i}X_{j} + \gamma\sum_{i \neq j} C_{i,j}
\end{equation}

The first and third terms of the fitness function are the benefits for a greater magnitude of opinion and more connection with neighboring nodes respectively and their forms are self-explanatory. For a given node, the second term sums over all other nodes in the network, determining the cognitive dissonance accrued by interaction with each of them~\cite{Akerlof:1982,Acharya:2016}. Note that where $X_{i}$ and $X_{j}$ are of opposite sign, meaning that the nodes $i$ and $j$ disagree, their contribution to the second term of the fitness function will be negative. Thus disagreeing nodes incur a penalty. The penalty is scaled by the strength of the connection between the two nodes, as the effects of cognitive dissonance will be most pronounced when there is a strong influence between individuals, thus making contradictory opinions harder to ignore. 

The purpose of our model is to understand the behavior of an opinion network given a set of initial conditions including the starting connections between nodes of the network, the initial beliefs of the nodes in the network, and the parameters of the fitness function. The coevolution of network connections and opinions is achieved through an iterative process of mutation (perturbation) that amounts to a simplified genetic algorithm: changes are made to both connection and opinion values in the network during each iteration, with predetermined step-sizes. At each step, only those changes that improve the fitness of the network are kept (see details in the Methods section). Over many iterations, these random mutations improve the network, leading ultimately to a resting stable state.

\section{Results}

Let us start with qualitatively understanding the long term coevolutionary dynamics of network connections and individual opinions. We make the assumption that the nodes' initial beliefs are distributed randomly over the full range of beliefs and that there exists no initial bias on the connections between node (nodes of similar opinions are not already clustered for instance). As shown in Eq~\eqref{eq:nodeFitness}, the fitness function of an individual $i$ can be  improved unilaterally by increasing the magnitude of its opinion $X_i$, regardless of their network connections. This suggests that individuals tend to adopt extreme opinions, either $+1$ or $-1$ in the long term, and the direction will be determined by the initial configurations within their local neighborhoods. 

Furthermore, taking the partial derivatives of $f_{i}$ with respect to $C_{ij}$, we obtain 
\begin{eqnarray}
\label{eq:partialDerivatives}
\frac{\partial f_{i}(X,C)}{\partial C_{i,j}} = \beta X_{i}X_{j} + \gamma.
\end{eqnarray}
We find that node $i$ is likely to increase the strength of its connection to node $j$, if $X_{i}X_{j} > -\gamma/\beta$. Similarly we find that node $i$ tends to increase its magnitude of opinion, if $\sum{X_{j}C_{i,j}} > - \sign(X_i)\alpha/\beta$. In particular, for $\beta > \gamma$, nodes with opposite views are much likely to become disconnected, and hence the resulting network tends to divide into with two echo chambers with splitting views. In order to illustrate these effects, we turn to computer simulations to study how the long-term behavior of the network depends on the parameters $\beta$ and $\gamma$ (see Methods). 

\begin{figure}[ht]
\centering
\includegraphics[width=\linewidth]{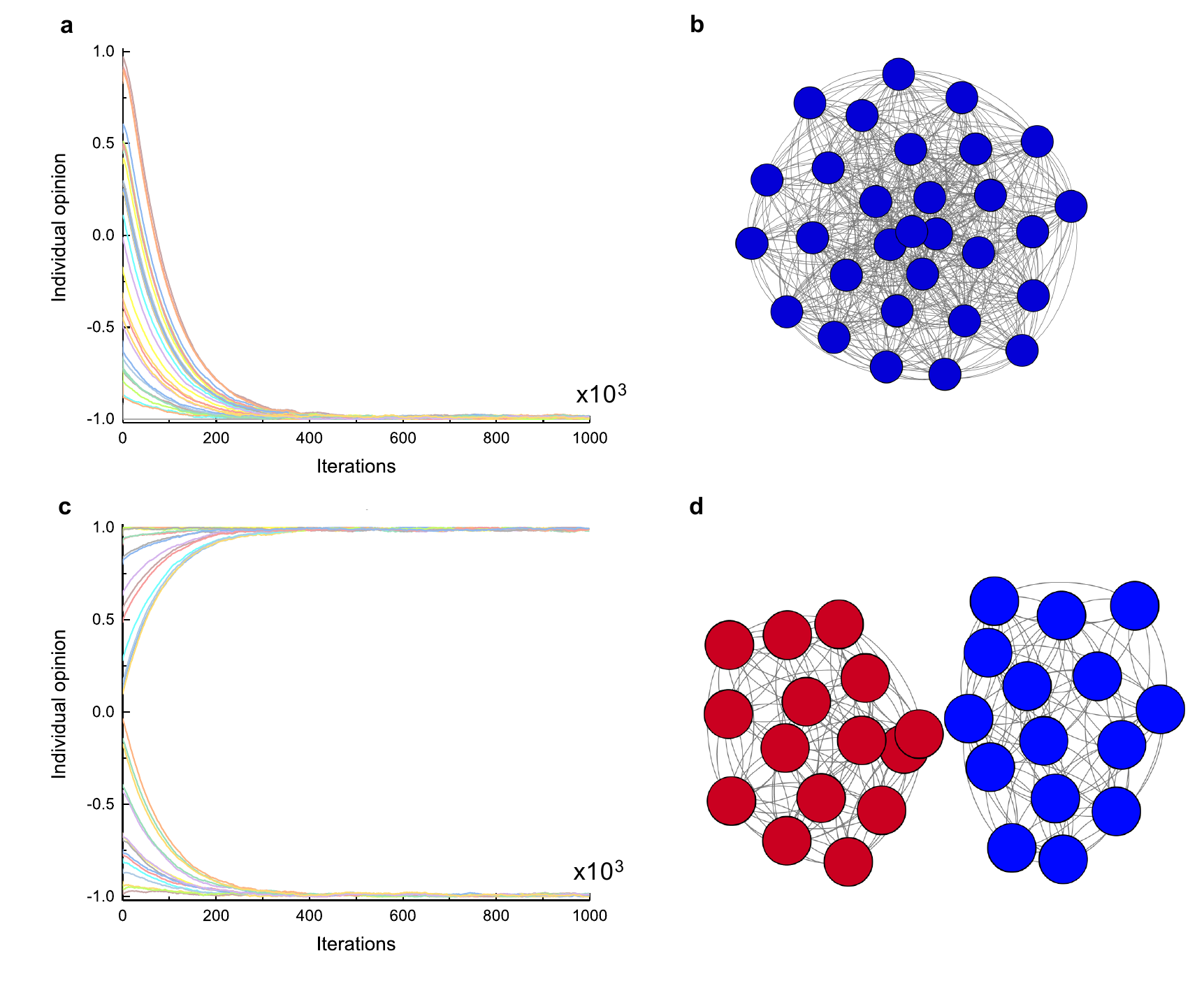}
\caption{Coevolution of opinions and networks, where individuals can adjust their opinions as well as their network connections. Panels (a) and (b) show an example of cohesive community with consensus in opinion (Parameters:\(\alpha = .5, \beta = .5, \gamma = .7\)). Panels (c) and (d) show the formation of echo chambers (Parameters: \(\alpha = .5, \beta = .05, \gamma = .7\)). }
\label{fig1}
\end{figure}

Figure 1 demonstrates that the parameter choices of $\beta$ and $\gamma$ mainly drive the coevolutionary dynamics of opinion and networks. Figure  \ref{fig1}a and \ref{fig1}b reflect what can happen when $\beta$ is dominant rather than $\gamma$. There is an initial stage of divergence, but in this case it reduces overtime. As the value of maintaining connections is lower than it was for the convergent network, the relationships between nodes that disagree quickly deteriorate, leaving entirely separate populations. In this final state, the nodes in the network have no cognitive dissonance. Their opinions are entirely aligned with those of the other nodes present in their local community. We can relate this scenario to the real-world phenomena of political echo chambers. People with similar opinions group together to the extent that they never hear opinions differing from their own.

Figure \ref{fig1}c and \ref{fig1}d show that cohesive communities with consensus of opinion can be obtained for fixed $\alpha$ and $\beta < \gamma$. In this case, there is often an initial phase in which opinions become more extreme on both sides of the argument, but then are turned around such that the network converges to consensus (Fig.~\ref{fig1}d). This apparent change in opinion is a result of the co-evolving network structure, which at the turn around point has created sufficient connections between individuals that do not agree, such that there is greater benefit to moving toward the same opinion than otherwise (Fig.~\ref{fig1}c). With $\gamma$ higher, however, the likelihood of connection overwhelming initial extremism is higher as early on, more connections are formed. We might expect such an arrangement in the real world in the case of family members: here it is difficult to rapidly change the level of communication with other individuals on a given topic but there is still benefit to maintaining cognitive consistency. We would expect this to promote the ultimate convergence of opinions in the local network at least.  

\begin{figure}[ht]
\centering
\includegraphics[width=\linewidth]{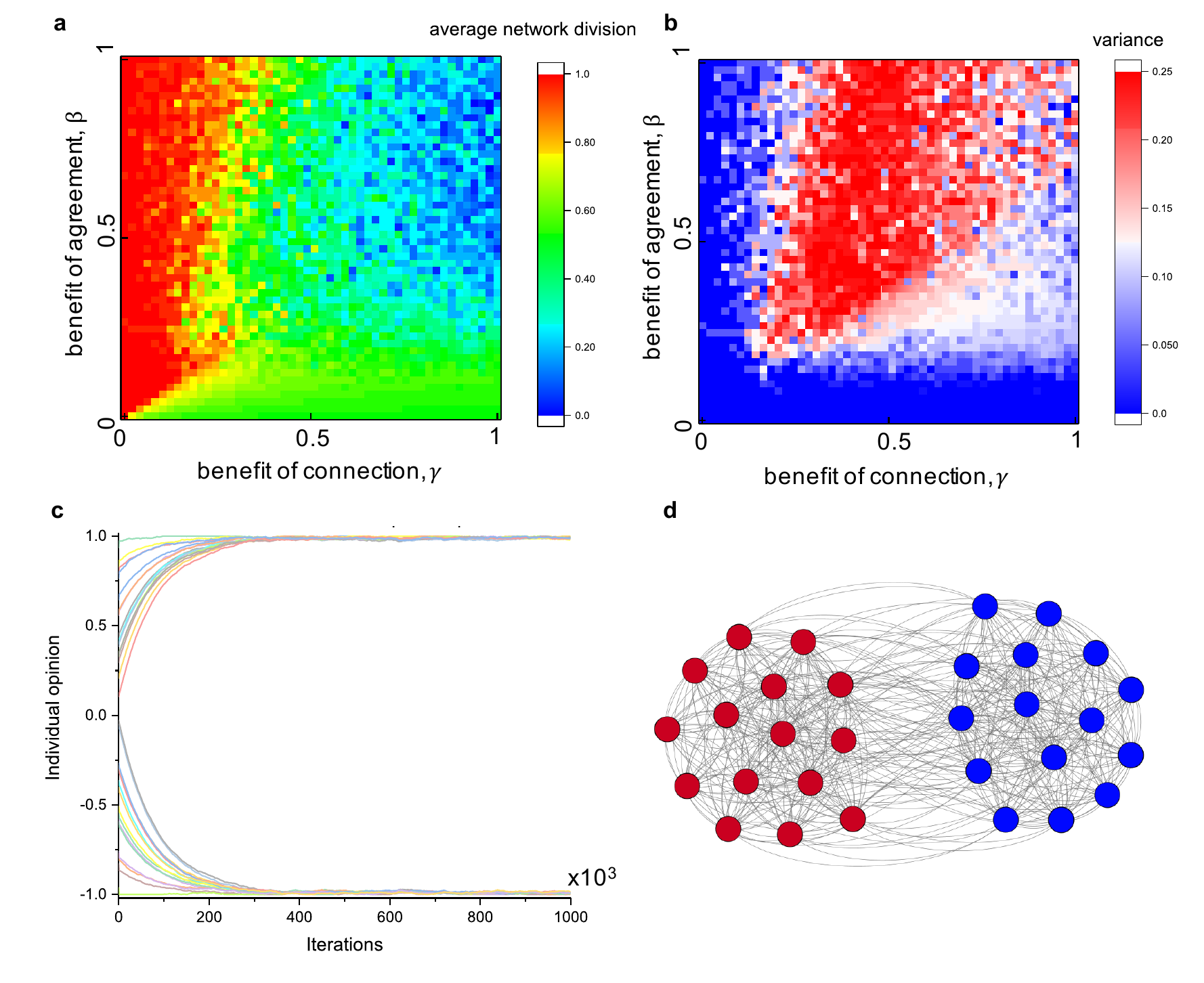}
\caption{Conditions for the emergence of echo chambers. (Parameters: \(\alpha = .5, \beta = .7, \gamma = .05\))}
\label{fig2}
\end{figure}

To help gain a complete picture of these results, we perform comprehensive simulations across the parameter space $(\gamma, \beta)$, as shown in Fig.~\ref{fig2}. This illustrates some of the intuition that might be gained from the analytical solutions for the behavior of a single node. Where $\beta$ exceeds $\gamma$, it is generally expected that there will be division for example, and this is indeed what we see in the plot (Fig.~\ref{fig2}a and \ref{fig2}b). Where $\beta$ is small, it is expected that there will be division of opinion, but with a reasonable level of connections in the network. These are the cases located near the bottom of the plot (Fig.~\ref{fig2}a). Interestingly, the divisions between these critical regions are not perfect; there are sprays of random model runs on both sides of the various divides in the plot which defy the expectation given the parameter values. Over many runs, we use these values to form a plot of the variance across the parameter space as in Fig.~\ref{fig2}b.

As revealed by our sensitivity analysis (Fig. 2a and 2b), besides entirely split networks and fully connected networks (Fig. 1), it is possible to have sub-community formation with multiple levels of inter-group connections (Fig. 2c and 2d). Indeed this is what we find in the case of the U.S House of Representatives voting records~\cite{Johnson:2012}: there is rarely complete consensus or complete division.

We can understand these simulation results intuitively as follows. The $\alpha$ and $\beta$ terms tend to drive the network opinions apart and the connections between them to zero, whereas $\gamma$ term tends to pull the network together. The specific parameter choices determine the balance between these two forces, and therefore can lead to a variety of typical behaviors of the co-evolutionary dynamics of opinions and networks (Fig. 1 and 2). It should be noted that the influence of the initial structure of the network on its final state may be drastic. There exist cases in which the initial structure of the network, due to the random initialization used, biases the behavior of the network to such a great extent that it reverses what would have happened in a graph without any such bias. In particular these might be cases where by chance a random division in the connections of the network is aligned with the divisions in the opinions of the nodes. This might lead to division in the network regardless of the ratio between the parameters $\beta$ and $\gamma$. Alternatively there may be a case in which there is strong initial connection between nodes of different opinions, promoting the development of consensus where they might normally have been division. 


There may also be initial biases, arising from the random initialization by chance, in the distributions of the opinions of the nodes. If the belief values chosen are particularly skewed to one side, even where the values are within their prescribed positive or negative ranges, it may be the case that the the nodes will reach consensus when the parameters would normally have induced division. Similarly, a distribution of beliefs that contains only extreme opinions to begin with may stay that way even if the parameters might have predicted consensus. It is notable that the model does not allow people to hold on to weak opinion values for very long. Even for weak $\alpha$ values, once groups are somewhat separated, the $\beta$ term leads to a mutual reinforcement of opinion between nodes. The greater the magnitudes of the beliefs of the nodes in question, the larger the $\beta$ term can become, benefiting both of the nodes in question.

As a real-world application of our model, we use voting records of the United States House of Representatives over a timespan of decades~\cite{Johnson:2012}, in order to understand the influence of underlying psychological and social factors on increasing partisanship in recent years. This dataset has been previously considered as an excellent example for the measurement of network division over time~\cite{Johnson:2012}. House members were considered to have a positive belief level if they identified with the Democrats and negative belief level if they were Republican. While party affiliation clearly encompasses a broad range of beliefs, we take it as a proxy for belief on a particular topic. 

\begin{figure}[ht]
\centering
\includegraphics[width=\linewidth]{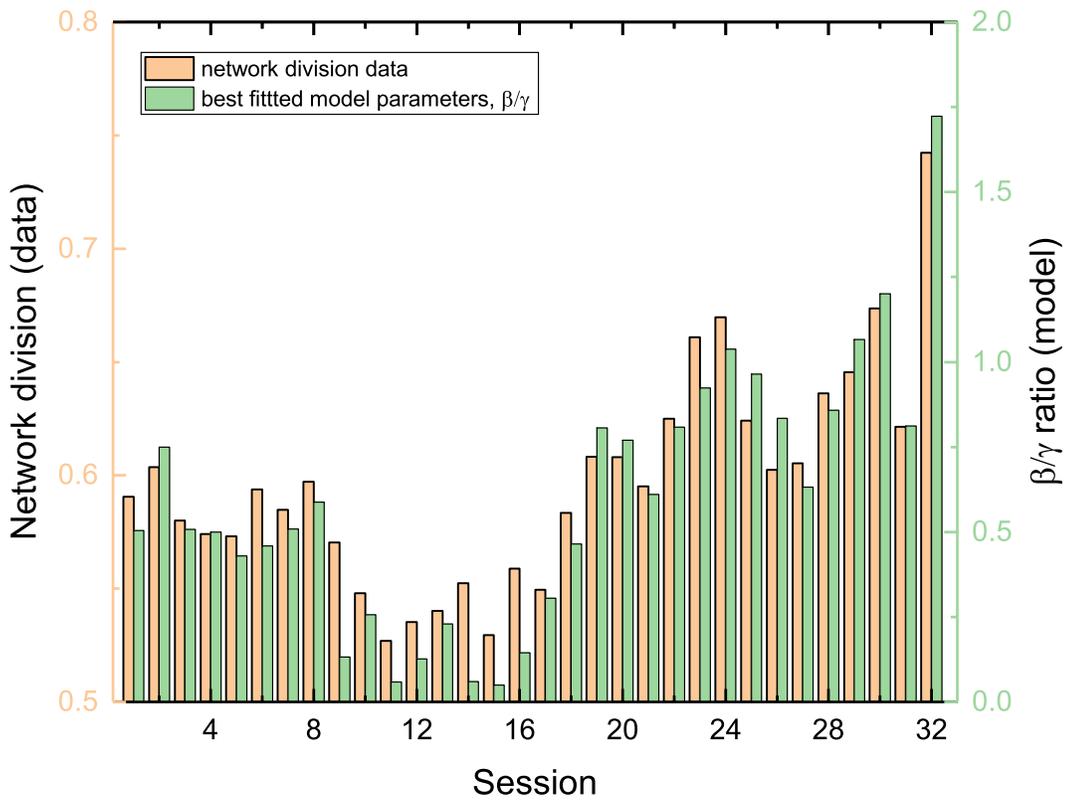}
\caption{Inferred model parameters for given network division. Parameters: \(\alpha = .5\)
}
\label{fig3}
\end{figure}

\begin{figure}[ht]
\centering
\includegraphics[width=\linewidth]{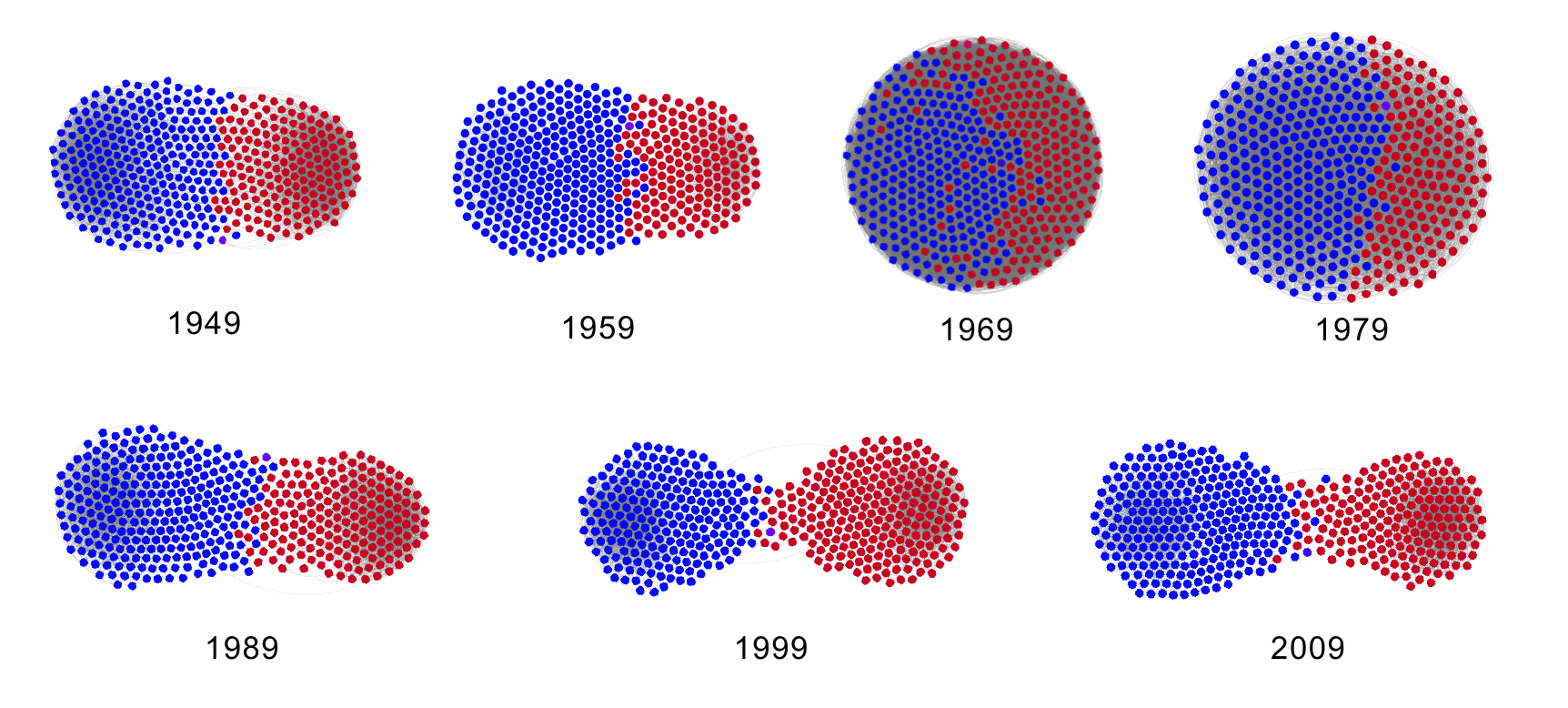}
\caption{Snapshots of House of Representatives networks generated using our model with inferred model parameter values $\beta$ and $\gamma$ (\(\alpha = .5\)). 
}
\label{fig4}
\end{figure}

Figure~\ref{fig3} shows the calculated division values for different trial years, which span over the last five decades (each year indicating a session of the House of Representatives), along with the inferred model parameter ratio $\beta/\gamma$. Figure~\ref{fig4} shows network snapshots over time that are generated by our model with best-fitted parameter values. We found evidence to support that the relative emphases of opinion conformity (represented by $\beta$) and general connection (represented by $\gamma$) have been changing over time. Over this time period, we can see that both network division values and the ratio of $\beta$ and $\gamma$ have first experienced decreases followed by dramatic increases. In particular, recent years see a clear growth in the level of division, which along with a commensurate growth in \(\beta/\gamma\), suggests a shift in priority from connection to congruity of opinion. These results demonstrate that our model and specifically the parameters of the fitness function can be used as a way to understand changes in opinion networks over time. Greater network division can be modeled as a growth in the value of local homogeneity of opinion as compared to the general value placed on connections with other nodes. 

\section{Discussion \& Conclusion}
Our model provides a method for the simulation of changing opinions on a network, allowing for the co-evolution of opinions and the connections between nodes. We have shown that it is possible to influence the final shape of a network through the modification of a fitness function based on the minimization of cognitive dissonance. Example runs of the model have demonstrated that the initial distribution of the nodes in the network and of their opinions can tip the balance of cases where the parameters of the fitness function does not strongly promote division or cohesion in the final network, but that the behavior of the networks can be connected to those parameter values in intuitive ways. 

The current model allows for the simultaneous evolution of both the connections between the nodes and their individual opinions. The purpose of the evolution is to maximize the fitness of the network, where the definition of fitness includes the effects of cognitive dissonance, node degree, and opinion magnitude. This is in line with other models that have considered the interplay between individual forces acting on a given node and the combined forces of other nodes in the network. \cite{Moon:2015} The resulting formulation is reminiscent of a spin-glass model, but is not equivalent (See Castellano for a review of spin-glass models and similar constructions \cite{Castellano:2009}). Cognitive dissonance is defined as the discomfort felt by an individual who holds two competing opinions at the same time. The discomfort it produces prompts individuals to avoid cases where they are confronted with ideas that do not match their own \cite{Akerlof:1982,Axelrod:1997,Acharya:2016,Gawronski:2017}. This brings out the primary assumption of this paper: people do not form their opinions based on evidence, but rather choose the opinions that they hold and the individuals with whom they surround themselves for the purpose of maximizing fitness as we have defined it.

An application of our model to partisanship in the United States House of Representatives suggests that it is possible to estimate the fitness function parameters for a real-world network and that the results are logically connected to the meaning of the different parameters. The case study points to a growth in the importance of homophily at the expense of having fewer connections in American politics. The result is greater partisanship and polarization, and has been widely documented \cite{Polarization:2014}. The fitness function forces changes in the network that ultimately lead to clustering of those nodes who share the same opinions and a division of the network into mutually disaffected populations. This structure mimics that of "echo chambers" in the media and in politics, and is self-reinforcing. If this trend continues, we are likely to see only greater political division in the future rather than a return toward greater cohesion. While some variety is necessary for effective problem-solving, it is concerning to think that politics may become so divided that no compromises can be reached on even the most important questions. Data sets pertaining to changing social networks and the flow of information through them have become significantly easier to access in recent decades with the advent of the Internet and large-scale social networking sites \cite{Newman:2003,Adamic:2003,Borner:2004}. The modeling of Twitter or Facebook networks as they change in time can be, for instance, another natural application of this research.

The importance of the underlying substrate topology has been demonstrated clearly for the prisoner's dilemma and similar games~\cite{Abramson:2001,Helbing:2010,Roca:2009}. We find that the same is true of opinion dynamics. Initial conditions can heavily influence the final behavior of a given network. One of the fundamental assumptions of this paper was that individuals choose their opinions and their connections only to avoid disagreement with their neighbors in the social network. This is indeed enough to create structure in the network similar to the structures that we see played out in real world social networks, but it does also beg the question of where the validity of a particular opinion actually comes into play. When does the evidence start to matter in the formation of opinions? Further research should consider the interplay between logical evidence and the influence of other individuals through the medium of cognitive dissonance. A model able to combine these two components effectively would be much more complete and more accurate in the prediction of future states of a given network of opinion. Future research will also consider the application of multiple levels of opinion, delving into the world of belief "systems" in contrast to single beliefs~\cite{Friedkin:2016}.

\section{Methods}

\subsection{Simulation procedure}
Summing over all of the nodes, we have $f(\mathbf{X},\mathbf{C}) = \sum_{i=1}^N f_{i}$ as the total fitness function for the entire network. Expanding this quantity we have:
\begin{eqnarray}
\label{eq:networkFitness}
f(\mathbf{X},\mathbf{C})  = \sum_{i=1}^N \Big(\alpha|X_{i}|+\beta\sum_{i \neq j}C_{i,j}X_{i}X_{j} + \gamma\sum_{i \neq j} C_{i,j}\Big)
\end{eqnarray}

The quantity $f(\mathbf{X},\mathbf{C})$ is used as a measure of the progress of the network as a whole toward the steady state solution. Any possible improvements to $f(\mathbf{X},\mathbf{C})$ eventually become more and more rare as the network approaches a steady state, and so the rate of actual change of $f$ with respect to each iteration of the model decreases. 

Specifically, the update step of the model consists of two rounds of random changes to the network, first to the connections and then to the beliefs of the nodes. The number of connections that are changed and the number of beliefs that are changed can be manipulated. The case for which only one node is changed at a time is similar to a Moran process \cite{Moran:1962}, and a change of all of the nodes during each step would be akin to a Wright-Fisher update process \cite{Tran:2012}. The changes in this model are not made with either of these update processes explicitly, however. Each round includes a validation stage which ensures that the changes made have improved the fitness of the nodes in the network. Where there is no change or negative change to the fitness of a node, the corresponding mutation is removed and the state of that node variable is reset to its value at the end of the last iteration. The maximum value of a mutation can also be manipulated, but remains fixed throughout the course of the model run. This is particularly useful for the exploration of edge cases where mutations of either the connections or the beliefs of the nodes approach zero, at which point the connections or beliefs are fixed.  

We take \(X_{i}(t)\) to denote the belief of the $i$th node at time $t$ and \(X'_{i}(t)\) as the state after mutation in time step $t$. By the same token \(C_{ij}\) and \(C'_{ij}\) are the values of the connection between the $i$th and $j$th nodes at time $t$ and the their mutated values. Let \(\Delta X_{i}\) indicate the mutation in the belief of the $i$th node while \(\Delta C_{ij}\) indicates the mutation of the connection between the $i$th and $j$th nodes. The value of this mutation is pulled from a uniform distribution in the range \([-.001,.001]\). We then have the following as the update rule for each of the nodes in the network:

\begin{eqnarray}
\label{eq:belMut}
X'_{i}(t) = X_{i}(t) + \Delta X_{i}
\end{eqnarray}
\begin{eqnarray}
\label{eq:belUpdate}
X_{t+1} := 
	\begin{cases} 
      X'_{i}(t) & f_{i}(X'(t),C(t)) \geq f_{i}(X(t),C(t))\\
      X_{i}(t) & f_{i}(X'(t),C(t)) < f_{i}(X(t),C(t))\\
   \end{cases}
\end{eqnarray}

Similarly for the connections between nodes we have the following update rule (with mutations in the range \([-.0001,.0001]\)):
\begin{eqnarray}
\label{eq:conMut}
C'_{ij}(t) = C_{ij}(t) + \Delta C_{ij}
\end{eqnarray}
\begin{eqnarray}
\label{eq:conUpdate}
C_{ij}(t+1) := 
	\begin{cases} 
      C'_{ij}(t) & f_{i}(X(t),C'(t)) \geq f_{i}(X(t),C(t))\\
      C_{ij}(t) & f_{i}(X(t),C'(t)) < f_{i}(X(t),C(t))\\
   \end{cases}
\end{eqnarray}

In MATLAB, the above update rules are enacted as matrix operations, with a vector of mutations being added to $\mathbf{X}$ and then validated, while a matrix of changes is added to $\mathbf{C}$ before validation. 

Eventually this process brings the network to a steady state solution where neither the opinions of the nodes nor the connections between them can be changed in a way that is beneficial to the nodes of the network. 

\subsection*{Network division}
In order to compare the final states of the networks produced by the model, we introduce a simple metric for network division, which we define as the degree to which two groups are disconnected in a network. The two groups in the case of opinion dynamics will be those who agree and those who disagree. The sums of the weights on connections between nodes of the same type (denoted $\mathbf{I}$) and the weight on connections between nodes of different types (denoted $\mathbf{O}$) are calculated. The value of network division is then calculated to be \(\frac{\mathbf{I}}{\mathbf{I}+\mathbf{O}}\). For convenience, we define networks where all nodes converge to the same belief value as having network division values of 0. Where there are only connections in the network between nodes of the same type, the value of network division will be 1. Partially split networks will have values less than 1, but greater than 0 where there are any in-group connections at all. Network division will be equal to zero in the case of a bipartite graph, where there are only intergroup connections.

\subsection*{Model-data integration}
In applying this model to data concerning voting in the U.S. House of Representatives, we calculate the division value using the average number of times that individuals voted together. The in-group connections for democrats becomes the number of democrats for a given year multiplied by the average number of democrat-democrat voting pairs, with a similar construction applied to the republican members. The total in-group connections $\mathbf{I}$ is defined as the sum of these two products. The out-group connections $\mathbf{O}$ is defined as the average number of democrat-republican pairs multiplied with the total number of members. A network division value for each session of the House of Representatives can then be calculated using the formula described above. 

As a simplifying assumption, we take the beliefs of the individuals of congress to be steady, and fix the value of $\alpha$ at $0.5$. Theoretically, the parameter $\alpha$ should have no effect in this scenario, as taking the partial derivative of the fitness function with respect to the connections between nodes entirely removes the first term. 

We wished to find the $\beta/\gamma$ values that would produce these same values of network division in a similar network. We begin with a population of nodes that had the same ratio of ``democrats'' to ``republicans'', meaning nodes with positive or negative belief values, as was found in a given session of the House of Representatives. We then ran the model on this network for 40 different ratios of $\beta/\gamma$ for each session. Performing a linear regression on the resultant set of data points, we gain a hypothesized relationship between the ratio $\beta/\gamma$ and network division for that particular arrangement of nodes. We can then use our hypothesized function to find what value of $\beta/\gamma$ would produce the network division observed. This same process is applied to all of the sessions for which there is data.

\section*{Acknowledgements}
We are grateful for support from the Dartmouth Faculty Startup Fund, Walter \& Constance Burke Research Initiation Award, and NIH Roybal Center Pilot Grant. T.E. acknowledges the support from Dartmouth Presidential Scholarship Program. 


\section*{Data, code and materials}
Code and data that can be used to replicate the results in the paper are available upon request.  


\begin{thebibliography}{47}
\expandafter\ifx\csname natexlab\endcsname\relax\def\natexlab#1{#1}\fi
\expandafter\ifx\csname bibnamefont\endcsname\relax
  \def\bibnamefont#1{#1}\fi
\expandafter\ifx\csname bibfnamefont\endcsname\relax
  \def\bibfnamefont#1{#1}\fi
\expandafter\ifx\csname citenamefont\endcsname\relax
  \def\citenamefont#1{#1}\fi
\expandafter\ifx\csname url\endcsname\relax
  \def\url#1{\texttt{#1}}\fi
\expandafter\ifx\csname urlprefix\endcsname\relax\def\urlprefix{URL }\fi
\providecommand{\bibinfo}[2]{#2}
\providecommand{\eprint}[2][]{\url{#2}}

\bibitem[{\citenamefont{Albert et~al.}(2000)\citenamefont{Albert, Jeong, and
  Barabási}}]{Albert:2000}
\bibinfo{author}{\bibfnamefont{R.}~\bibnamefont{Albert}},
  \bibinfo{author}{\bibfnamefont{H.}~\bibnamefont{Jeong}}, \bibnamefont{and}
  \bibinfo{author}{\bibfnamefont{A.-L.} \bibnamefont{Barabási}},
  \bibinfo{journal}{Nature} \textbf{\bibinfo{volume}{406}},
  \bibinfo{pages}{378} (\bibinfo{year}{2000}).

\bibitem[{\citenamefont{Milgram}(1967)}]{Milgram:1967}
\bibinfo{author}{\bibfnamefont{S.}~\bibnamefont{Milgram}},
  \bibinfo{journal}{Psychology Today} \textbf{\bibinfo{volume}{2}},
  \bibinfo{pages}{60} (\bibinfo{year}{1967}).

\bibitem[{\citenamefont{Borgatti}(2003)}]{Borgatti:2003}
\bibinfo{author}{\bibfnamefont{S.~P.} \bibnamefont{Borgatti}},
  \emph{\bibinfo{title}{Dynamic Social Network Modeling and Analysis: Workshop
  Summary and Papers.}} (\bibinfo{publisher}{Committee on Human Factors,
  National Research Council}, \bibinfo{year}{2003}), chap.
  \bibinfo{chapter}{The Key Player Problem}, pp. \bibinfo{pages}{241--252}.

\bibitem[{\citenamefont{Newman}(2003)}]{Newman:2003}
\bibinfo{author}{\bibfnamefont{M.~E.~J.} \bibnamefont{Newman}},
  \bibinfo{journal}{SIAM Review} \textbf{\bibinfo{volume}{45}},
  \bibinfo{pages}{167} (\bibinfo{year}{2003}).

\bibitem[{\citenamefont{Newman}(2006)}]{Newman:2006}
\bibinfo{author}{\bibfnamefont{M.~E.~J.} \bibnamefont{Newman}},
  \bibinfo{journal}{PNAS} \textbf{\bibinfo{volume}{103}}, \bibinfo{pages}{8577}
  (\bibinfo{year}{2006}).

\bibitem[{\citenamefont{M{\"a}s et~al.}(2010)\citenamefont{M{\"a}s, Flache, and
  Helbing}}]{Mas:2010}
\bibinfo{author}{\bibfnamefont{M.}~\bibnamefont{M{\"a}s}},
  \bibinfo{author}{\bibfnamefont{A.}~\bibnamefont{Flache}}, \bibnamefont{and}
  \bibinfo{author}{\bibfnamefont{D.}~\bibnamefont{Helbing}},
  \bibinfo{journal}{Computational Biology} \textbf{\bibinfo{volume}{6}}
  (\bibinfo{year}{2010}).

\bibitem[{\citenamefont{Zadorozhnyi and Yudin}(2012)}]{Zadorozhnyi:2012}
\bibinfo{author}{\bibfnamefont{V.}~\bibnamefont{Zadorozhnyi}} \bibnamefont{and}
  \bibinfo{author}{\bibfnamefont{E.}~\bibnamefont{Yudin}},
  \bibinfo{journal}{Automation and Control} \textbf{\bibinfo{volume}{73}},
  \bibinfo{pages}{702} (\bibinfo{year}{2012}).

\bibitem[{\citenamefont{J{\'a}come et~al.}(2010)\citenamefont{J{\'a}come,
  da~Silva, Moreira, Jr., and Herrmann}}]{Jacome:2010}
\bibinfo{author}{\bibfnamefont{S.}~\bibnamefont{J{\'a}come}},
  \bibinfo{author}{\bibfnamefont{L.}~\bibnamefont{da~Silva}},
  \bibinfo{author}{\bibfnamefont{A.}~\bibnamefont{Moreira}},
  \bibinfo{author}{\bibfnamefont{J.~A.} \bibnamefont{Jr.}}, \bibnamefont{and}
  \bibinfo{author}{\bibfnamefont{H.}~\bibnamefont{Herrmann}},
  \bibinfo{journal}{Physica A}  (\bibinfo{year}{2010}).

\bibitem[{\citenamefont{Dorogovtsev et~al.}(2008)\citenamefont{Dorogovtsev,
  Goltsev, and Mendes}}]{Dorogovtsev:2008}
\bibinfo{author}{\bibfnamefont{S.~N.} \bibnamefont{Dorogovtsev}},
  \bibinfo{author}{\bibfnamefont{A.~V.} \bibnamefont{Goltsev}},
  \bibnamefont{and} \bibinfo{author}{\bibfnamefont{J.~F.~F.}
  \bibnamefont{Mendes}}, \bibinfo{journal}{Reviews of Modern Physics}
  \textbf{\bibinfo{volume}{80}} (\bibinfo{year}{2008}).

\bibitem[{\citenamefont{Dorogovtsev and Mendes}(2002)}]{Dorogovtsev:2002}
\bibinfo{author}{\bibfnamefont{S.~N.} \bibnamefont{Dorogovtsev}}
  \bibnamefont{and} \bibinfo{author}{\bibfnamefont{J.~F.~F.}
  \bibnamefont{Mendes}}, \bibinfo{journal}{Advances in Physics}
  \textbf{\bibinfo{volume}{51}} (\bibinfo{year}{2002}).

\bibitem[{\citenamefont{Erd{\"o}s and Renyi}(1959)}]{Erdos:1959}
\bibinfo{author}{\bibfnamefont{P.}~\bibnamefont{Erd{\"o}s}} \bibnamefont{and}
  \bibinfo{author}{\bibfnamefont{A.}~\bibnamefont{Renyi}},
  \bibinfo{journal}{Publicationes Mathimaticae} \textbf{\bibinfo{volume}{6}},
  \bibinfo{pages}{290} (\bibinfo{year}{1959}).

\bibitem[{\citenamefont{Erd{\"o}s and Renyi}(1960)}]{Erdos:1960}
\bibinfo{author}{\bibfnamefont{P.}~\bibnamefont{Erd{\"o}s}} \bibnamefont{and}
  \bibinfo{author}{\bibfnamefont{A.}~\bibnamefont{Renyi}},
  \bibinfo{journal}{Magyar Tudom{\'a}nyos Akad{\'e}mia Matematikai Kutat{\'o}
  Int{\'e}zet{\'e}nek K{\"o}zlem{\'e}nyei} \textbf{\bibinfo{volume}{5}},
  \bibinfo{pages}{17} (\bibinfo{year}{1960}).

\bibitem[{\citenamefont{Sznajd-Weron and Sznajd}(2000)}]{sznajd2000opinion}
\bibinfo{author}{\bibfnamefont{K.}~\bibnamefont{Sznajd-Weron}}
  \bibnamefont{and} \bibinfo{author}{\bibfnamefont{J.}~\bibnamefont{Sznajd}},
  \bibinfo{journal}{International Journal of Modern Physics C}
  \textbf{\bibinfo{volume}{11}}, \bibinfo{pages}{1157} (\bibinfo{year}{2000}).

\bibitem[{\citenamefont{Deffuant et~al.}(2002)\citenamefont{Deffuant, Amblard,
  Weisbuch, and Faure}}]{deffuant2002can}
\bibinfo{author}{\bibfnamefont{G.}~\bibnamefont{Deffuant}},
  \bibinfo{author}{\bibfnamefont{F.}~\bibnamefont{Amblard}},
  \bibinfo{author}{\bibfnamefont{G.}~\bibnamefont{Weisbuch}}, \bibnamefont{and}
  \bibinfo{author}{\bibfnamefont{T.}~\bibnamefont{Faure}},
  \bibinfo{journal}{Journal of artificial societies and social simulation}
  \textbf{\bibinfo{volume}{5}} (\bibinfo{year}{2002}).

\bibitem[{\citenamefont{Sznajd-Weron}(2005)}]{sznajd2005sznajd}
\bibinfo{author}{\bibfnamefont{K.}~\bibnamefont{Sznajd-Weron}},
  \bibinfo{journal}{arXiv preprint physics/0503239}  (\bibinfo{year}{2005}).

\bibitem[{\citenamefont{Watts and Dodds}(2007)}]{watts2007influentials}
\bibinfo{author}{\bibfnamefont{D.~J.} \bibnamefont{Watts}} \bibnamefont{and}
  \bibinfo{author}{\bibfnamefont{P.~S.} \bibnamefont{Dodds}},
  \bibinfo{journal}{Journal of consumer research}
  \textbf{\bibinfo{volume}{34}}, \bibinfo{pages}{441} (\bibinfo{year}{2007}).

\bibitem[{\citenamefont{Szab{\'o} and Szolnoki}(2002)}]{szabo2002three}
\bibinfo{author}{\bibfnamefont{G.}~\bibnamefont{Szab{\'o}}} \bibnamefont{and}
  \bibinfo{author}{\bibfnamefont{A.}~\bibnamefont{Szolnoki}},
  \bibinfo{journal}{Physical Review E} \textbf{\bibinfo{volume}{65}},
  \bibinfo{pages}{036115} (\bibinfo{year}{2002}).

\bibitem[{\citenamefont{Sood et~al.}(2008)\citenamefont{Sood, Antal, and
  Redner}}]{sood2008voter}
\bibinfo{author}{\bibfnamefont{V.}~\bibnamefont{Sood}},
  \bibinfo{author}{\bibfnamefont{T.}~\bibnamefont{Antal}}, \bibnamefont{and}
  \bibinfo{author}{\bibfnamefont{S.}~\bibnamefont{Redner}},
  \bibinfo{journal}{Physical Review E} \textbf{\bibinfo{volume}{77}},
  \bibinfo{pages}{041121} (\bibinfo{year}{2008}).

\bibitem[{\citenamefont{Redner}(2017)}]{redner2017dynamics}
\bibinfo{author}{\bibfnamefont{S.}~\bibnamefont{Redner}},
  \bibinfo{journal}{arXiv preprint arXiv:1705.02249}  (\bibinfo{year}{2017}).

\bibitem[{\citenamefont{Jalili and Perc}(2017)}]{jalili2017information}
\bibinfo{author}{\bibfnamefont{M.}~\bibnamefont{Jalili}} \bibnamefont{and}
  \bibinfo{author}{\bibfnamefont{M.}~\bibnamefont{Perc}},
  \bibinfo{journal}{Journal of Complex Networks} \textbf{\bibinfo{volume}{5}},
  \bibinfo{pages}{665} (\bibinfo{year}{2017}).

\bibitem[{\citenamefont{Friedkin et~al.}(2016)\citenamefont{Friedkin,
  Proskurnikov, Tempo, and Parsegov}}]{Friedkin:2016}
\bibinfo{author}{\bibfnamefont{N.~E.} \bibnamefont{Friedkin}},
  \bibinfo{author}{\bibfnamefont{A.~V.} \bibnamefont{Proskurnikov}},
  \bibinfo{author}{\bibfnamefont{R.}~\bibnamefont{Tempo}}, \bibnamefont{and}
  \bibinfo{author}{\bibfnamefont{S.~E.} \bibnamefont{Parsegov}},
  \bibinfo{journal}{Science} \textbf{\bibinfo{volume}{354(6310)}},
  \bibinfo{pages}{321} (\bibinfo{year}{2016}).

\bibitem[{\citenamefont{Axelrod}(1997)}]{Axelrod:1997}
\bibinfo{author}{\bibfnamefont{R.}~\bibnamefont{Axelrod}},
  \bibinfo{journal}{The Journal of Conflict Resolution}
  \textbf{\bibinfo{volume}{41}}, \bibinfo{pages}{203} (\bibinfo{year}{1997}).

\bibitem[{\citenamefont{Holme and Newman}(2006)}]{holme2006nonequilibrium}
\bibinfo{author}{\bibfnamefont{P.}~\bibnamefont{Holme}} \bibnamefont{and}
  \bibinfo{author}{\bibfnamefont{M.~E.} \bibnamefont{Newman}},
  \bibinfo{journal}{Physical Review E} \textbf{\bibinfo{volume}{74}},
  \bibinfo{pages}{056108} (\bibinfo{year}{2006}).

\bibitem[{\citenamefont{Zanette and Gil}(2006)}]{zanette2006opinion}
\bibinfo{author}{\bibfnamefont{D.~H.} \bibnamefont{Zanette}} \bibnamefont{and}
  \bibinfo{author}{\bibfnamefont{S.}~\bibnamefont{Gil}},
  \bibinfo{journal}{Physica D: Nonlinear Phenomena}
  \textbf{\bibinfo{volume}{224}}, \bibinfo{pages}{156} (\bibinfo{year}{2006}).

\bibitem[{\citenamefont{Nardini et~al.}(2008)\citenamefont{Nardini, Kozma, and
  Barrat}}]{nardini2008s}
\bibinfo{author}{\bibfnamefont{C.}~\bibnamefont{Nardini}},
  \bibinfo{author}{\bibfnamefont{B.}~\bibnamefont{Kozma}}, \bibnamefont{and}
  \bibinfo{author}{\bibfnamefont{A.}~\bibnamefont{Barrat}},
  \bibinfo{journal}{Physical review letters} \textbf{\bibinfo{volume}{100}},
  \bibinfo{pages}{158701} (\bibinfo{year}{2008}).

\bibitem[{\citenamefont{Fu and Wang}(2008)}]{Fu:2008}
\bibinfo{author}{\bibfnamefont{F.}~\bibnamefont{Fu}} \bibnamefont{and}
  \bibinfo{author}{\bibfnamefont{L.}~\bibnamefont{Wang}},
  \bibinfo{journal}{Phys. Rev. E} \textbf{\bibinfo{volume}{78}}
  (\bibinfo{year}{2008}),
  \urlprefix\url{https://link.aps.org/doi/10.1103/PhysRevE.78.016104}.

\bibitem[{\citenamefont{Perc and Szolnoki}(2010)}]{perc2010coevolutionary}
\bibinfo{author}{\bibfnamefont{M.}~\bibnamefont{Perc}} \bibnamefont{and}
  \bibinfo{author}{\bibfnamefont{A.}~\bibnamefont{Szolnoki}},
  \bibinfo{journal}{BioSystems} \textbf{\bibinfo{volume}{99}},
  \bibinfo{pages}{109} (\bibinfo{year}{2010}).

\bibitem[{Pol(2014)}]{Polarization:2014}
\bibinfo{type}{Tech. Rep.}, \bibinfo{institution}{Pew Research Center}
  (\bibinfo{year}{2014}).

\bibitem[{Par(2016)}]{Partisanship:2016}
\bibinfo{type}{Tech. Rep.}, \bibinfo{institution}{Pew Research Center}
  (\bibinfo{year}{2016}).

\bibitem[{\citenamefont{{Schmidt} et~al.}(2018)\citenamefont{{Schmidt},
  {Zollo}, {Scala}, {Betsch}, and {Quattrociocchi}}}]{Schmidt:2018}
\bibinfo{author}{\bibfnamefont{A.~L.} \bibnamefont{{Schmidt}}},
  \bibinfo{author}{\bibfnamefont{F.}~\bibnamefont{{Zollo}}},
  \bibinfo{author}{\bibfnamefont{A.}~\bibnamefont{{Scala}}},
  \bibinfo{author}{\bibfnamefont{C.}~\bibnamefont{{Betsch}}}, \bibnamefont{and}
  \bibinfo{author}{\bibfnamefont{W.}~\bibnamefont{{Quattrociocchi}}},
  \bibinfo{journal}{ArXiv e-prints}  (\bibinfo{year}{2018}),
  \eprint{1801.02903}.

\bibitem[{\citenamefont{Andris et~al.}(2015)\citenamefont{Andris, Lee,
  Hamilton, Martino, Gunning, and Selden}}]{Andris:2015}
\bibinfo{author}{\bibfnamefont{C.}~\bibnamefont{Andris}},
  \bibinfo{author}{\bibfnamefont{D.}~\bibnamefont{Lee}},
  \bibinfo{author}{\bibfnamefont{M.}~\bibnamefont{Hamilton}},
  \bibinfo{author}{\bibfnamefont{M.}~\bibnamefont{Martino}},
  \bibinfo{author}{\bibfnamefont{C.}~\bibnamefont{Gunning}}, \bibnamefont{and}
  \bibinfo{author}{\bibfnamefont{J.}~\bibnamefont{Selden}},
  \bibinfo{journal}{PLoS ONE} \textbf{\bibinfo{volume}{10}}
  (\bibinfo{year}{2015}).

\bibitem[{Con(2017)}]{Conflict:2017}
\bibinfo{type}{Tech. Rep.}, \bibinfo{institution}{Pew Research Center}
  (\bibinfo{year}{2017}).

\bibitem[{Ele(2016)}]{Election:2016}
\bibinfo{type}{Tech. Rep.}, \bibinfo{institution}{Pew Research Center}
  (\bibinfo{year}{2016}).

\bibitem[{\citenamefont{Antonopoulos and Shang}(2018)}]{Antonopoulos:2017}
\bibinfo{author}{\bibfnamefont{C.~G.} \bibnamefont{Antonopoulos}}
  \bibnamefont{and} \bibinfo{author}{\bibfnamefont{Y.}~\bibnamefont{Shang}},
  \bibinfo{journal}{Scientific Reports}  (\bibinfo{year}{2018}).

\bibitem[{\citenamefont{Akerlof and Dickens}(Jun., 1982)}]{Akerlof:1982}
\bibinfo{author}{\bibfnamefont{G.~A.} \bibnamefont{Akerlof}} \bibnamefont{and}
  \bibinfo{author}{\bibfnamefont{W.~T.} \bibnamefont{Dickens}},
  \bibinfo{journal}{The American Economic Review}
  \textbf{\bibinfo{volume}{72(3)}}, \bibinfo{pages}{307} (\bibinfo{year}{Jun.,
  1982}).

\bibitem[{\citenamefont{Acharya et~al.}(2016)\citenamefont{Acharya, Blackwell,
  and Sen}}]{Acharya:2016}
\bibinfo{author}{\bibfnamefont{A.}~\bibnamefont{Acharya}},
  \bibinfo{author}{\bibfnamefont{M.}~\bibnamefont{Blackwell}},
  \bibnamefont{and} \bibinfo{author}{\bibfnamefont{M.}~\bibnamefont{Sen}},
  \emph{\bibinfo{title}{Explaining preferences from behavior: A cognitive
  dissonance approach}} (\bibinfo{year}{2016}).

\bibitem[{\citenamefont{{Johnson} et~al.}(2017)\citenamefont{{Johnson},
  {Manrique}, {Zheng}, {Cao}, {Botero}, {Huang}, {Aden}, {Song}, {Leady},
  {Velasquez} et~al.}}]{Johnson:2012}
\bibinfo{author}{\bibfnamefont{N.~F.} \bibnamefont{{Johnson}}},
  \bibinfo{author}{\bibfnamefont{P.}~\bibnamefont{{Manrique}}},
  \bibinfo{author}{\bibfnamefont{M.}~\bibnamefont{{Zheng}}},
  \bibinfo{author}{\bibfnamefont{Z.}~\bibnamefont{{Cao}}},
  \bibinfo{author}{\bibfnamefont{J.}~\bibnamefont{{Botero}}},
  \bibinfo{author}{\bibfnamefont{S.}~\bibnamefont{{Huang}}},
  \bibinfo{author}{\bibfnamefont{N.}~\bibnamefont{{Aden}}},
  \bibinfo{author}{\bibfnamefont{C.}~\bibnamefont{{Song}}},
  \bibinfo{author}{\bibfnamefont{J.}~\bibnamefont{{Leady}}},
  \bibinfo{author}{\bibfnamefont{N.}~\bibnamefont{{Velasquez}}},
  \bibnamefont{et~al.}, \bibinfo{journal}{ArXiv e-prints}
  (\bibinfo{year}{2017}), \eprint{1712.06009}.

\bibitem[{\citenamefont{Moon and Lu}(2015)}]{Moon:2015}
\bibinfo{author}{\bibfnamefont{H.}~\bibnamefont{Moon}} \bibnamefont{and}
  \bibinfo{author}{\bibfnamefont{T.-C.} \bibnamefont{Lu}},
  \bibinfo{journal}{Scientific Reports} \textbf{\bibinfo{volume}{5}}
  (\bibinfo{year}{2015}).

\bibitem[{\citenamefont{Castellano et~al.}(2009)\citenamefont{Castellano,
  Fortunato, and Loreto}}]{Castellano:2009}
\bibinfo{author}{\bibfnamefont{C.}~\bibnamefont{Castellano}},
  \bibinfo{author}{\bibfnamefont{S.}~\bibnamefont{Fortunato}},
  \bibnamefont{and} \bibinfo{author}{\bibfnamefont{V.}~\bibnamefont{Loreto}},
  \bibinfo{journal}{Rev. Mod. Phys.} \textbf{\bibinfo{volume}{81:2}}
  (\bibinfo{year}{2009}).

\bibitem[{\citenamefont{Gawronski and Brannon}(2017)}]{Gawronski:2017}
\bibinfo{author}{\bibfnamefont{B.}~\bibnamefont{Gawronski}} \bibnamefont{and}
  \bibinfo{author}{\bibfnamefont{S.~M.} \bibnamefont{Brannon}},
  \emph{\bibinfo{title}{Cognitive dissonance:Progress on a pivotal theory in
  social psychology}} (\bibinfo{publisher}{American Psychological Association},
  \bibinfo{year}{2017}), chap. \bibinfo{chapter}{What is Cognitive Consistency
  and Why Does it Matter?}, \bibinfo{edition}{2nd} ed.

\bibitem[{\citenamefont{Adamic and Adar}(2003)}]{Adamic:2003}
\bibinfo{author}{\bibfnamefont{L.~A.} \bibnamefont{Adamic}} \bibnamefont{and}
  \bibinfo{author}{\bibfnamefont{E.}~\bibnamefont{Adar}},
  \bibinfo{journal}{Social Networks} \textbf{\bibinfo{volume}{25}},
  \bibinfo{pages}{211} (\bibinfo{year}{2003}).

\bibitem[{\citenamefont{Borner et~al.}(2004)\citenamefont{Borner, Maru, and
  Goldstone}}]{Borner:2004}
\bibinfo{author}{\bibfnamefont{K.}~\bibnamefont{Borner}},
  \bibinfo{author}{\bibfnamefont{J.~T.} \bibnamefont{Maru}}, \bibnamefont{and}
  \bibinfo{author}{\bibfnamefont{R.~L.} \bibnamefont{Goldstone}},
  \bibinfo{journal}{PNAS} \textbf{\bibinfo{volume}{101}}, \bibinfo{pages}{5266}
  (\bibinfo{year}{2004}).

\bibitem[{\citenamefont{Guillermo~Abramson}(2001)}]{Abramson:2001}
\bibinfo{author}{\bibfnamefont{M.~K.} \bibnamefont{Guillermo~Abramson}},
  \bibinfo{journal}{Physical Review E} \textbf{\bibinfo{volume}{63}}
  (\bibinfo{year}{2001}).

\bibitem[{\citenamefont{Helbing and Lozano}(2010)}]{Helbing:2010}
\bibinfo{author}{\bibfnamefont{D.}~\bibnamefont{Helbing}} \bibnamefont{and}
  \bibinfo{author}{\bibfnamefont{S.}~\bibnamefont{Lozano}},
  \bibinfo{journal}{Physical Review E} \textbf{\bibinfo{volume}{81}}
  (\bibinfo{year}{2010}).

\bibitem[{\citenamefont{Roca et~al.}(2009)\citenamefont{Roca, Cuesta, and
  S{\'a}nchez}}]{Roca:2009}
\bibinfo{author}{\bibfnamefont{C.~P.} \bibnamefont{Roca}},
  \bibinfo{author}{\bibfnamefont{J.~A.} \bibnamefont{Cuesta}},
  \bibnamefont{and}
  \bibinfo{author}{\bibfnamefont{A.}~\bibnamefont{S{\'a}nchez}},
  \bibinfo{journal}{Science Direct}  (\bibinfo{year}{2009}).

\bibitem[{\citenamefont{Moran}(1962)}]{Moran:1962}
\bibinfo{author}{\bibfnamefont{P.}~\bibnamefont{Moran}},
  \emph{\bibinfo{title}{The statistical processes of evolutionary theory}}
  (\bibinfo{year}{1962}), chap. \bibinfo{chapter}{The statistical processes of
  evolutionary theory}.

\bibitem[{\citenamefont{Tran et~al.}(2012)\citenamefont{Tran, Hofrichter, and
  Jost}}]{Tran:2012}
\bibinfo{author}{\bibfnamefont{T.~D.} \bibnamefont{Tran}},
  \bibinfo{author}{\bibfnamefont{J.}~\bibnamefont{Hofrichter}},
  \bibnamefont{and} \bibinfo{author}{\bibfnamefont{J.}~\bibnamefont{Jost}},
  \bibinfo{journal}{Theory Biosci.} pp. \bibinfo{pages}{73--82}
  (\bibinfo{year}{2012}).

\end{thebibliography}

\clearpage

\clearpage

\clearpage

\clearpage

\end{document}